\documentclass[fleqn,usenatbib,useAMS]{mnras}
\pdfoutput=1

\usepackage{newtxtext,newtxmath}

\usepackage[T1]{fontenc}

\DeclareRobustCommand{\VAN}[3]{#2}
\let\VANthebibliography\thebibliography
\def\thebibliography{\DeclareRobustCommand{\VAN}[3]{##3}\VANthebibliography}

\usepackage{graphicx}	
\usepackage{amsmath}	
\usepackage{amssymb}	
\usepackage{soul}
\newcommand{\tco}{$^{13}$CO}
\newcommand{\cdo}{C$^{18}$O}
\newcommand{\kms}{\mbox{km~s$^{-1}$}}
\newcommand{\tyso}{\mbox{HH\ 900} YSO} 
\newcommand{\Msun}{\mbox{$\text{M}_{\sun}$}}

\title[Environmental impact]{Illuminating a tadpole's metamorphosis III: quantifying past and present environmental impact}

\author[M. Reiter et al.]{Megan Reiter,$^{1}$\thanks{E-mail: megan.reiter@stfc.ac.uk (MR)}
Thomas J. Haworth,$^{2}$
Andr{\'e}s E. Guzm{\'a}n,$^{3}$
Pamela D. Klaassen,$^{1}$
\newauthor Anna F. McLeod,$^{4,5}$
Guido Garay$^{6}$
\\
$^{1}$UK Astronomy Technology Centre, Blackford Hill, Edinburgh, EH9 3HJ, UK \\
$^{2}$Astronomy Unit, School of Physics and Astronomy, Queen Mary University of London, London E1 4NS, UK \\
$^{3}$National Astronomical Observatory of Japan, National Institutes of Natural Sciences, 2-21-1 Osawa, Mitaka, Tokyo 181-8588, Japan \\
$^{4}$Department of Astronomy, University of California Berkeley, Berkeley, CA 94720, USA\\
$^{5}$Department of Physics and Astronomy, Texas Tech University, PO Box 41051, Lubbock, TX 79409, USA\\
$^{6}$Departamento de Astronom\'{i}a, Universidad de Chile, Camino el Observatorio 1515, Las Condes, Santiago, Chile 
}

\date{Accepted XXX. Received YYY; in original form ZZZ}

\pubyear{2015}

\begin{document}
\label{firstpage}
\pagerange{\pageref{firstpage}--\pageref{lastpage}}
\maketitle

\begin{abstract}
We combine MUSE and ALMA observations with theoretical models to evaluate how a tadpole-shaped globule located in the Carina Nebula has been influenced by its environment. This globule is now relatively small (radius $\sim 2500$~au), hosts a protostellar jet+outflow (HH~900) and, with a blue-shifted velocity of  $\sim 10$~\kms, is travelling faster than it should be if its kinematics were set by the turbulent velocity dispersion of the precursor cloud.  Its outer layers are currently still subject to heating, but comparing the internal and external pressures implies that the globule is in a post-collapse phase. Intriguingly the outflow is bent, implying that the YSO responsible for launching it is comoving with the globule, which requires that the star formed after the globule was up to speed since otherwise it would have been left behind. We conclude that the most likely scenario is one in which the cloud was much larger before being subject to radiatively-driven implosion, which accelerated the globule to the high observed speeds under the photoevaporative rocket effect and triggered the formation of the star responsible for the outflow. The globule may now be in a quasi-steady state following collapse. Finally, the HH~900 YSO is likely $\gtrsim 1$~\Msun\ and may be the only star forming in the globule. It may be that this process of triggered star formation has prevented the globule from fragmenting to form multiple stars (e.g., due to heating) and has produced a single higher mass star. 
\end{abstract}




\section{Introduction}

 Most star formation happens in large aggregates that form both low- and high-mass stars. 
 In these regions, feedback from the high-mass stars  heats the gas \citep[e.g.,][]{guzman2015,marsh2017,lee2020}, may accelerate the destruction of protoplanetary disks \citep[e.g.,][]{mann2010,mann2014,eisner2018}, and ultimately destroys the star-forming cloud \citep[e.g.,][]{matzner2002,murray2010,chevance2019}.

 Stellar feedback clearly affects the surrounding gas, but how it impacts current and subsequent star formation is less straightforward. 
 Ionizing radiation and supernova shocks may provide an impulse to otherwise stable cores, driving them to collapse \citep[e.g.,][]{boss2008,gritschneder2012,li2014}. Most observational studies of triggered star formation rely on the spatial coincidence of dense gas, young stars, and a nearby high-mass star or collection of stars, although this does not demonstrate a causal connection \citep[see, e.g.,][]{desai2010,dale2015}.

 Multiple theoretical models have examined the mechanics of triggered star formation \citep[e.g.,][]{sandford1982,bertoldi1989,lefloch1994,esquivel2007,gritschneder2009,miao2009,2010MNRAS.403..714M, bisbas2011,haworth2012_diffuse,haworth2012_diagnostics,haworth2013}. 
 Radiatively-driven implosion (RDI) models typically start with a stable Bonnor-Ebert sphere that is illuminated with radiation of different intensities. 
 Historically, cloud morphology and mass-loss rates provided the best observational tests of simulations as cold gas and dust in individual cores was difficult or impossible to resolve at the typical distances of high-mass star-forming regions ($\gtrsim 2$~kpc).

 Bright-rimmed clouds have long been observed in H~{\sc ii} regions \citep[e.g.,][]{bok1948,pottasch1956,dyson1968,reipurth1983,1997A&A...324..249L, 2004A&A...415..627T, 2004A&A...428..723U, smith2004_finger,gahm2007,2008A&A...477..557M, 2009MNRAS.400.1726M,gahm2013}. 
Narrowband images help constrain the physical properties of these objects, especially the nature of the ionization front. 
In some cases, mass estimates have also been derived from these images, although these are often uncertain and differ substantially from masses derived from molecular line observations \citep[e.g.,][]{gahm2007,gahm2013,haikala2017,reiter2015_hh900,reiter2019_tadpole,reiter2020_tadpole}. 
Millimeter line observations from single-dish telescopes (where beamsizes are often an order of magnitude larger than the size of these small clouds)  
provide demographic differences between globules in different regions \citep[e.g.,][]{haikala2017}, but cannot constrain the physical properties in the cold molecular gas of the unresolved clouds.

With the advent of integral field unit spectrographs like the Multi-Unit Spectroscopic Explorer \citep[MUSE,][]{bacon2010} and sensitive interferometers like the Atacama Large Millimeter/sub-millimeter Array (ALMA), resolved observations of star-forming clouds subject to feedback from nearby high-mass stars is now a reality. 
In this paper, we consider recent spatially-resolved observations of a small (radius $\sim 1\arcsec$), tadpole-shaped globule in the Carina Nebula, which is shown in Figure \ref{fig:hh900_intro}. 
Ionizing radiation from nearby Tr16 slowly evaporates the system \citep[][hereafter Paper~I]{reiter2019_tadpole} while illuminating the globule and the HH~900 jet+outflow system that emerges from it \citep{smith2010,reiter2015_hh900}. 
The jet+outflow demonstrates that the globule is star-forming, although its density is so high that the jet-driving source is only seen with ALMA \citep[][hereafter Paper~II]{reiter2020_tadpole}. 
The dynamical age of the jet is $\lesssim 5000$~yr, indicating that the protostar is much younger than the nearby Tr16, where the most massive stars have already evolved off the main sequence \citep[suggesting an age $>3$~Myr, see e.g.,][]{walborn1995,smith2006_energy}. 
This hints that star formation in the globule happened recently, possibly under the influence of feedback \citep[e.g.,][]{bertoldi1989}. 

In this paper (Paper~III), we combine observational diagnostics from MUSE and ALMA (Papers~I and II, respectively) to understand how the external environment has affected the formation and evolution of the system.

\section{Data}

\subsection{MUSE}
Observations with the MUSE visual wavelength panoramic integral-field spectrograph on the VLT using the MUSE+GALACSI Adaptive Optics (AO) module in Wide Field Mode (WFM) were obtained on 03 April 2018.
The AO-assisted observations provide $\sim 0.8''$ imaging over the $1' \times 1'$ field-of-view. 
MUSE spectra cover a nominal wavelength range of 
$4650-9300$~\AA\ with a gap between $\sim 5800-5950$~\AA\ due to the laser guide stars for the AO system and provide 
spectral resolution of R=2000-4000, corresponding to a velocity resolution of $\sim$75--150~\kms.
A single-pointing with MUSE covers the full extent of the tadpole globule and HH~900 jet+outflow system. 
The total on-source integration time was $720$~s. 
A full description of the MUSE data may be found in Paper~I. 

\subsection{ALMA}
ALMA Band 7 and 6 observations of the HH~900 globule were obtained in 2016 and 2017, respectively. 
The spectral setup targeted the J=3-2 rotational transitions of $^{13}$CO and C$^{18}$O, as well as the J=2-1 transitions of 
$^{12}$CO, $^{13}$CO, and C$^{18}$O. 
Velocity resolution of all observed lines is $\lesssim 0.1$~\kms, with resolution ranging between $0.06$--$0.16$~\kms. 
The synthesized beamsizes of the reduced data range typically between 0\farcs1--0\farcs2, corresponding to a spatial resolution 230--460~au at the distance of Carina \citep[2.3~kpc;][]{smith2006_distance}.
A more complete description of the observations and data reduction can be found in Paper~II. 

\begin{figure*}
    \includegraphics[trim=0mm 0mm 0mm 0mm, width=17cm]{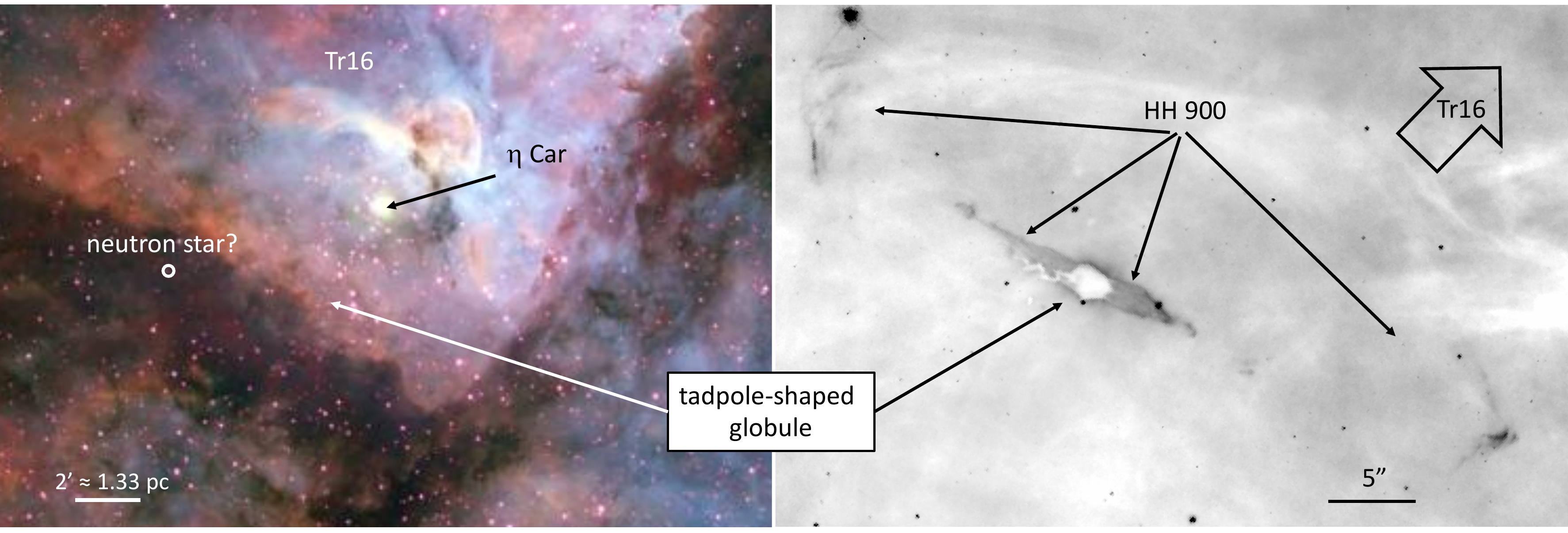}
    \caption{
    \textit{Left:} The location of the tadpole-shaped globule within the context of the Carina Nebula shown on a narrowband optical image 
    ([O~{\sc iii}]=blue, H$\alpha$=green, [S~{\sc ii}]=red; North is up, East is to the left; image credit: Nathan Smith, University of Minnesota/NOAO/AURA/NSF). 
    The Tr16 star cluster is located near the top of the image and includes the evolved star $\eta$~Carinae. We also note the location of the candidate neutron star identified by \citet{hamaguchi2009}. 
    \textit{Right:} A zoomed-in view of the tadpole-shaped globule and the HH~900 jet+outflow system shown on an H$\alpha$ image from \emph{HST}. 
    }
    \label{fig:hh900_intro}
\end{figure*}

\subsection{The distance to Carina}\label{ss:distance}
Several distance estimates for Carina exist in the literature. 
Uneven reddening over the region has been a persistent challenge for photometric distance estimates. 
More direct measures, like parallax data from the \emph{Gaia} mission, can help in this regard. 
Despite these challenges, many, though not all, distance estimates are consistent within the uncertainties.

In this paper, we adopt the estimate from \citet{smith2006_distance} who derived a distance of $2.3\pm0.05$~kpc from the kinematics of the expanding Homunculus nebula surrounding $\eta$~Carinae.  
Other methods suggest slightly larger distances. 
\citet{hur2012} estimate a distance of $2.9\pm 0.3$~kpc from main sequence fitting 
while  
\citet{povich2019} use \emph{Gaia} parallaxes to derive  $2.5^{+0.28}_{-0.23}$~kpc.  
We adopt 2.3~kpc as the most conservative choice. 
If the true distance is larger than this value, the estimated size of the globule will increase by $\sim 10-25$\% and the estimated density will decrease by $\sim 15-25$\%.

\section{Internal kinematics}
\subsection{Evidence of on-going environmental impact}\label{ss:environ}

In projection, the tadpole globule appears to lie 2.8~pc from Tr16. 
The globule is seen in silhouette in optical images, 
suggesting that it may lie in front of the bulk of the bright ionized nebulosity. 
If this is the case, the globule may be relatively unaffected by the environment. 
However, in Paper~I, we used the physical properties of the ionization front to estimate an incident ionizing flux 
of $\log(Q_H) \sim 48.3$~s$^{-1}$, similar to the ionizing photon flux from a late O-type star, suggesting Tr16 affects the globule. 
Additional evidence comes from the HH~900 jet+outflow system associated with the globule. 
The ionization in the jet+outflow increases with distance from the globule; that is, the longer the jet+outflow is in the H~{\sc ii} region, the more highly ionized it becomes. 
This suggests that the environment is ionizing both the jet+outflow and the globule.

\citet{bertoldi1989} explored how the stability of initially neutral clouds depends on the column density and the strength of the incident radiation. 
The physical properties of the ionized gas measured in Paper~I suggest an ionization front on the globule surface that is strong enough to compress the cloud for all mass estimates of the globule \citep[which range from $14$~M$_{\mathrm{Jupiter}}$ to $\sim 3.9$~\Msun; these all fall in region~II in Figure~1 of][]{bertoldi1989}. 
The tadpole-shaped globule is clearly star-forming, although whether its formation was triggered by the external feedback is difficult to prove \citep[e.g.,][]{dale2015}.

Nevertheless, the environment clearly continues to affect the physical properties in the globule. 
Data presented in Paper~II suggests that the globule has a positive radial temperature gradient with hotter gas near the surface of the globule while gas deep in the interior remains cold ($T<30$~K). 
Three lines of evidence suggest this temperature structure: 
(1) limb-brightened optically thick $^{12}$CO emission that is interpreted as higher temperatures near the surface of the globule; 
(2) a relatively flat spectral energy distribution of the millimeter continuum emission detected near the \tyso\ that is best fit with models that have lower dust temperatures ($T \lesssim 30$~K) and low $\beta$ (for a dust absorption coefficient that behaves as $\kappa_\nu=\kappa_0(\nu/\nu_0)^\beta$); and 
(3) the serendipitous detection of a deuterated species, DCN J=3-2, in the immediate environs of the \tyso. 
Deuterated species typically trace cold gas, where CO has frozen out of the gas phase \citep[e.g.,][]{bergin2007,caselli2012}.

\subsubsection{Comparing pressures}
To test whether external feedback is currently compressing the globule, we compare the present-day pressure of the ionization front with the internal support of the globule \citep[as in][]{smith2004_finger}.
Using the physical properties in the ionization front derived in Paper~I, we compute the pressure in the ionization front using the expression  
\begin{equation}
P_{IF} = n_e k T_e + m_H n_e v^2 
\end{equation}\label{eq:IF_pressure}
where
$n_e \approx 200$~cm$^{-3}$ is the electron density, 
$k$ is the Boltzmann constant, 
$T_e \approx 10^4$~K is the electron temperature, 
$m_H$ is the mass of hydrogen, and 
$v \approx 11$~km~s$^{-1}$ is the flow speed, assumed to be the sound speed of ionized gas. 
This expression includes both thermal pressure (first term) and back pressure from the photoevaporative flow (second term) which are of similar magnitude \citep[as noted by][]{smith2004_finger}. 

The pressure inside the neutral globule is given by 
\begin{equation}\label{eq:neutral_pressure}
P_0 = \frac{\rho_0}{\mu m_H}k T_{\mathrm{cold}} + \rho_0\sigma^2 + \frac{B^2}{8 \pi} = \rho_0 \left[ \left(\frac{\sigma_{\mathrm{thermal}}}{2}\right)^2 + \sigma_{\mathrm{turb}}^2 \right] + \frac{B^2}{8 \pi}  
\end{equation}
where
$\rho_0$ is the mass density, 
$\mu=2.37$ is the mean molecular weight, 
$\sigma$ is the velocity dispersion, and 
$B$ is the strength of the magnetic field.
Equation~\ref{eq:neutral_pressure} 
represents the summation of the thermal, turbulent, and magnetic pressures.  
We estimate the mass density and turbulent velocity dispersion from the C$^{18}$O which appears to be the least optically thick isotopologue and unaffected by the outflow. 
In Paper~II, we found the median C$^{18}$O linewidth (FWHM) to be $\Delta v = 1.4$~\kms, corresponding to $\sigma_{\mathrm{C^{18}O}} = 0.59$~\kms\ (using $\Delta v = 2 \sqrt{2 \ln(2)} \,\, \sigma$). 
Assuming a temperature $T_{\mathrm{cold}}=30$~K for the cold, molecular gas in the globule, we estimate the thermal velocity dispersion 
as $\sigma_{\mathrm{C^{18}O,thermal}} = \sqrt{2kT/ m_{\mathrm{C^{18}O}}} = 0.13$~\kms. 
To compute the turbulent velocity dispersion, we subtract the thermal component from the total dispersion, 
$\sigma_{\mathrm{turb, 1D}}^2 = \sigma_{\mathrm{C^{18}O}}^2 - \sigma_{\mathrm{C^{18}O, thermal}}^2$. 
We convert this 1D estimate to a 3D velocity dispersion using 
$\sigma_{\mathrm{turb}} = \sqrt{3} \sigma_{1D} \approx 1.0$~km~s$^{-1}$.

The thermal velocity dispersion of the molecular gas in the globule is 
$\sigma_{\mathrm{thermal, 1D}} = \sqrt{2 k_B T_{\mathrm{cold}} / \mu m_H} = 0.46$~\kms. 
This is similar to the turbulent velocity dispersion we compute from \cdo, $\sigma_{\mathrm{turb}, 1D} = 0.58$~\kms, suggesting that they contribute roughly equally to the support of the globule.

Neglecting magnetic pressure, we find $P_{IF}/P_0 \approx 0.009$ which indicates that the pressure of the ionization front does not currently drive the dynamics of the tadpole.
Including a non-zero magnetic field will increase the pressure in the neutral gas. 
While the high ionizing flux incident on the system suggests it will be susceptible to radiatively-driven collapse (see Section~\ref{ss:environ}), the high density and dominant neutral gas pressure suggest that collapse due to external pressure may have already happened. 

\citet{smith2004_finger} performed a similar analysis for another globule in Carina, the so-called Defiant Finger. 
The pressures in the Finger appear better matched with $P_{IF}/P_0 \approx 0.5$, leading \citet{smith2004_finger} to suggest that RDI may compress the globule and trigger star formation in the future.
Compared to the tadpole globule, the electron density in the ionization front is 20$\times$ higher in the Finger \citep{smith2004_finger,mcleod2016} while the turbulent velocity dispersion, estimated from single-dish (unresolved) molecular line observations from \citet{cox1995}, is 1.1~km~s$^{-1}$, remarkably similar to the tadpole (1.0~km~s$^{-1}$). 
\citet{smith2004_finger} estimate that the density of the cold molecular gas in the Finger is two orders of magnitude lower than we find in the tadpole (Paper~II). 
Unlike the tadpole, the Finger shows no sign of star formation.

\subsubsection{Estimating the critical ionizing flux}\label{sss:ionizing_flux}

\citet{bisbas2011} explored triggered star formation in clouds exposed to an external source of ionizing radiation. 
Higher fluxes produce ionization fronts that rapidly penetrate clouds, reducing the fraction of cloud mass converted to stars and the mass of stars formed. 
Above a critical threshold, $\Phi_{\mathrm{crit}} \gtrsim \Phi_{\mathrm{LyC}}$, clouds are dispersed without forming stars at all. 
\citet{bisbas2011} define the critical ionizing flux (their equation~10) as 
\begin{equation}
    \Phi_{\mathrm{crit}} = 6 \times 10^{13} cm^{-2} s^{-1} \left( \frac{M_{cloud}}{M_{\odot}} \right)^{-3}
    \label{equn:phicrit}
\end{equation}
where $M_{cloud}$ is the cloud mass. 
Using the globule mass from Paper~II, $M_{cloud} \approx 1.9$~\Msun, we find 
$\Phi_{\mathrm{crit}} \approx 8.7 \times 10^{12}$~cm$^{-2}$~s$^{-1}$.

In Paper~I, we estimated that the flux of ionizing photons that reaches the surface of the globule is $\log(Q_H) \sim 48.3$~s$^{-1}$, corresponding to 
$\Phi_{\mathrm{LyC}} \approx 2.1 \times 10^{9}$~cm$^{-2}$~s$^{-1}$ (assuming that the center of Tr16 is at a projected distance $D=2.8$~pc). 
This is more than three orders of magnitude below the critical flux.

We note that $\eta$~Carinae dominated the ionizing photon luminosity of Tr16 until recently, alone providing $\log(Q_H) \sim 50.77$~s$^{-1}$ from a projected distance $D=3.5$~pc. 
This corresponds to $\Phi_{\mathrm{LyC}} \approx 4.1 \times 10^{11}$~cm$^{-2}$~s$^{-1}$, a factor of $\sim 20$ lower than $\Phi_{\mathrm{crit}}$. 
Even this more extreme estimate places the globule firmly in the region of parameter space susceptible to star formation. 
Uncertainties in the distance between the globule and the ionizing sources as well as changes in the ionizing flux from Tr16 (especially as the high-mass stars like $\eta$~Car evolve off the main sequence) may produce different incident fluxes as a function of time.

\subsubsection{Cold gas kinematics}\label{sss:cold_kin}

The incident ionizing flux seen by the globule is below $\Phi_{\mathrm{crit}}$, indicating favorable environmental conditions for star formation. 
However, the current pressure inside the high-density globule ($n \gtrsim 10^6$~cm$^{-3}$) far exceeds that of the ionization front. 
The average density in the globule is more than two orders of magnitude larger than the initial central density ($\sim 10^{3.5}$~cm$^{-3}$) of the Bonnor-Ebert sphere in the simulations of \citet{bisbas2011}. 
Such circumstantial evidence hints that the globule has \emph{already} been compressed. 
Global infall, seen in the cold gas kinematics, would provide more compelling evidence. 

Self-absorbed molecular line profiles seen in other sources, particularly those that display a blue asymmetry, 
have been interpreted as evidence of infall  \citep[e.g.,][]{walker1986,tafalla1998,difrancesco2001,narayanan2002}. 
This is not observed in the tadpole; line profiles of all of the CO isotopologues tend to be single-peaked. 
In Paper~II, we argue that a positive radial thermal profile can explain the absence of self-absorbed profiles as hot intervening material will not absorb emission from the colder inner regions, and emission closer to the line peak traces warmer, more external gas. 
\begin{figure*}
    \includegraphics[trim=0mm 0mm 0mm 10mm,width=9cm]{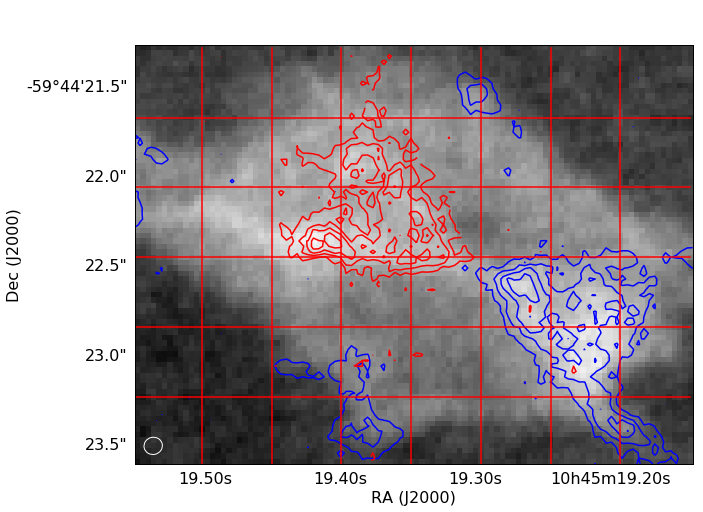} \\
    \includegraphics[trim=0mm 0mm 0mm 0mm, width=7.75cm]{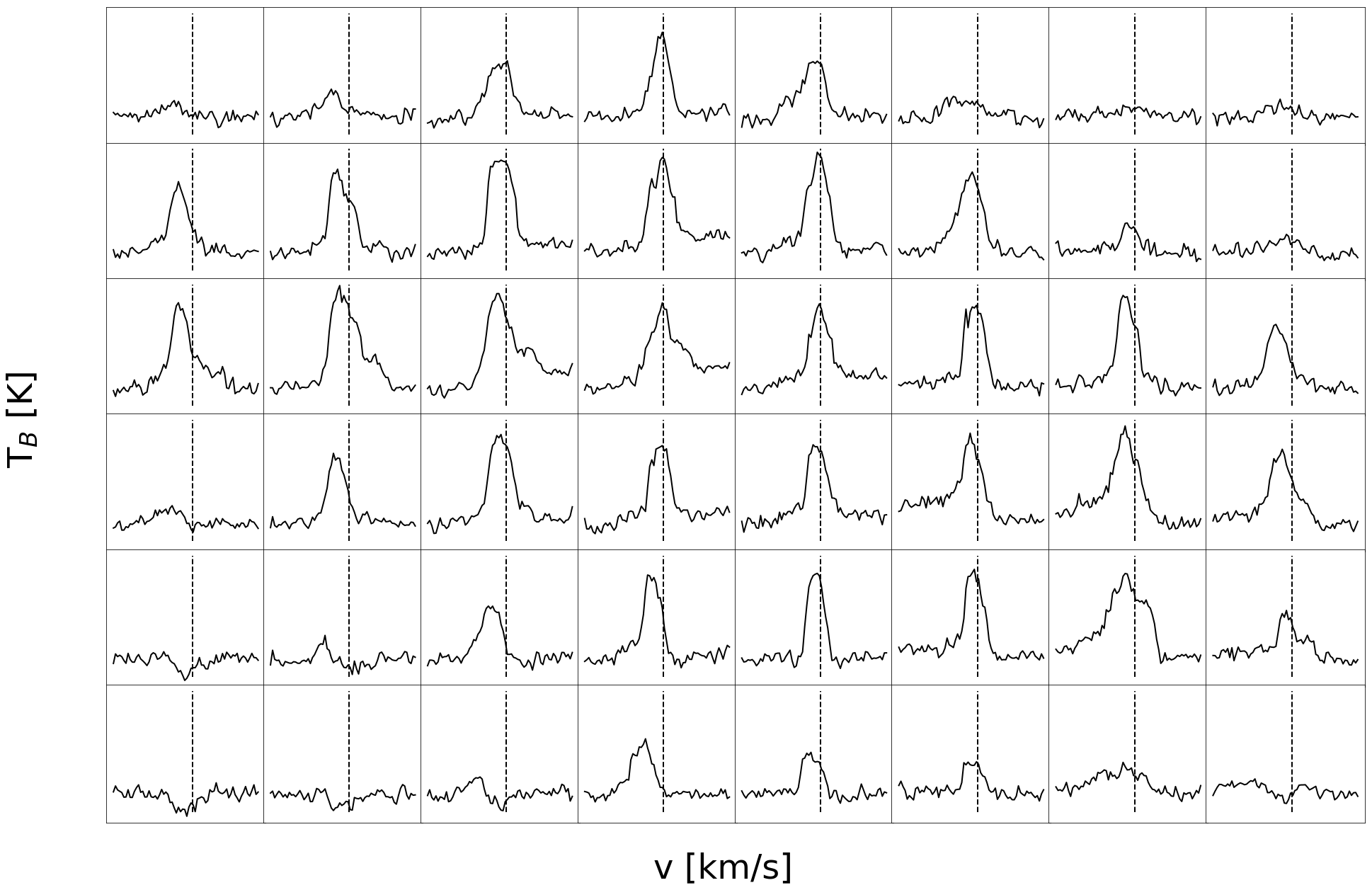}
    \includegraphics[trim=-10mm 0mm 0mm 5mm, width=8.5cm]{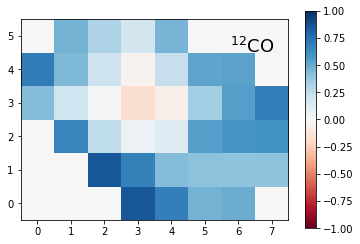}
    \includegraphics[trim=0mm 0mm 0mm 5mm,width=7.75cm]{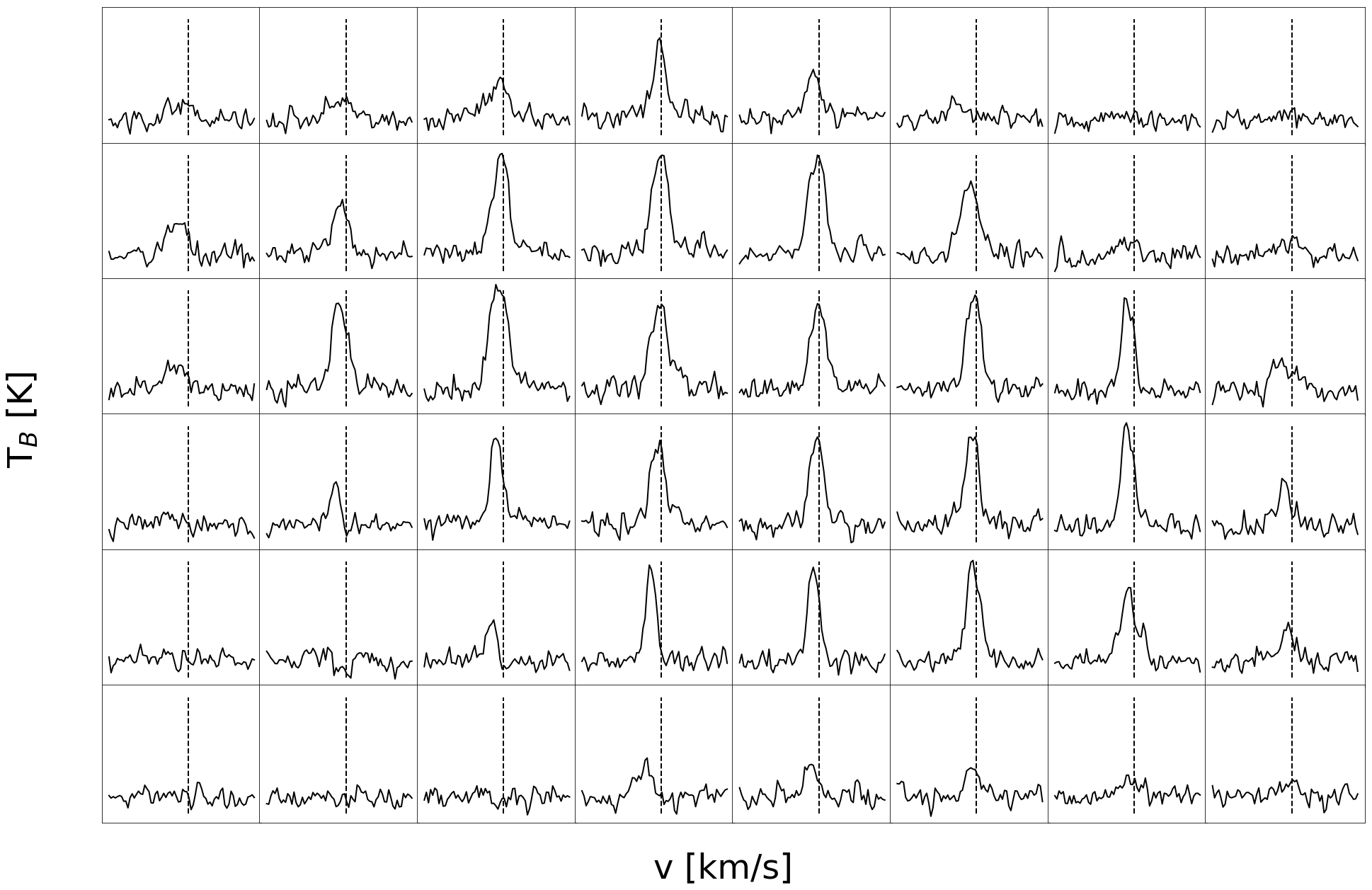}
    \includegraphics[trim=-10mm 0mm 0mm 5mm, width=8.5cm]{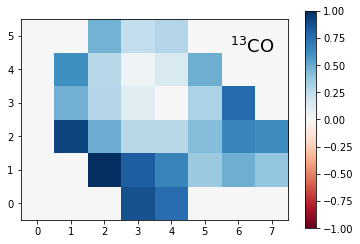}
    \includegraphics[trim=0mm 0mm 0mm 5mm,width=7.75cm]{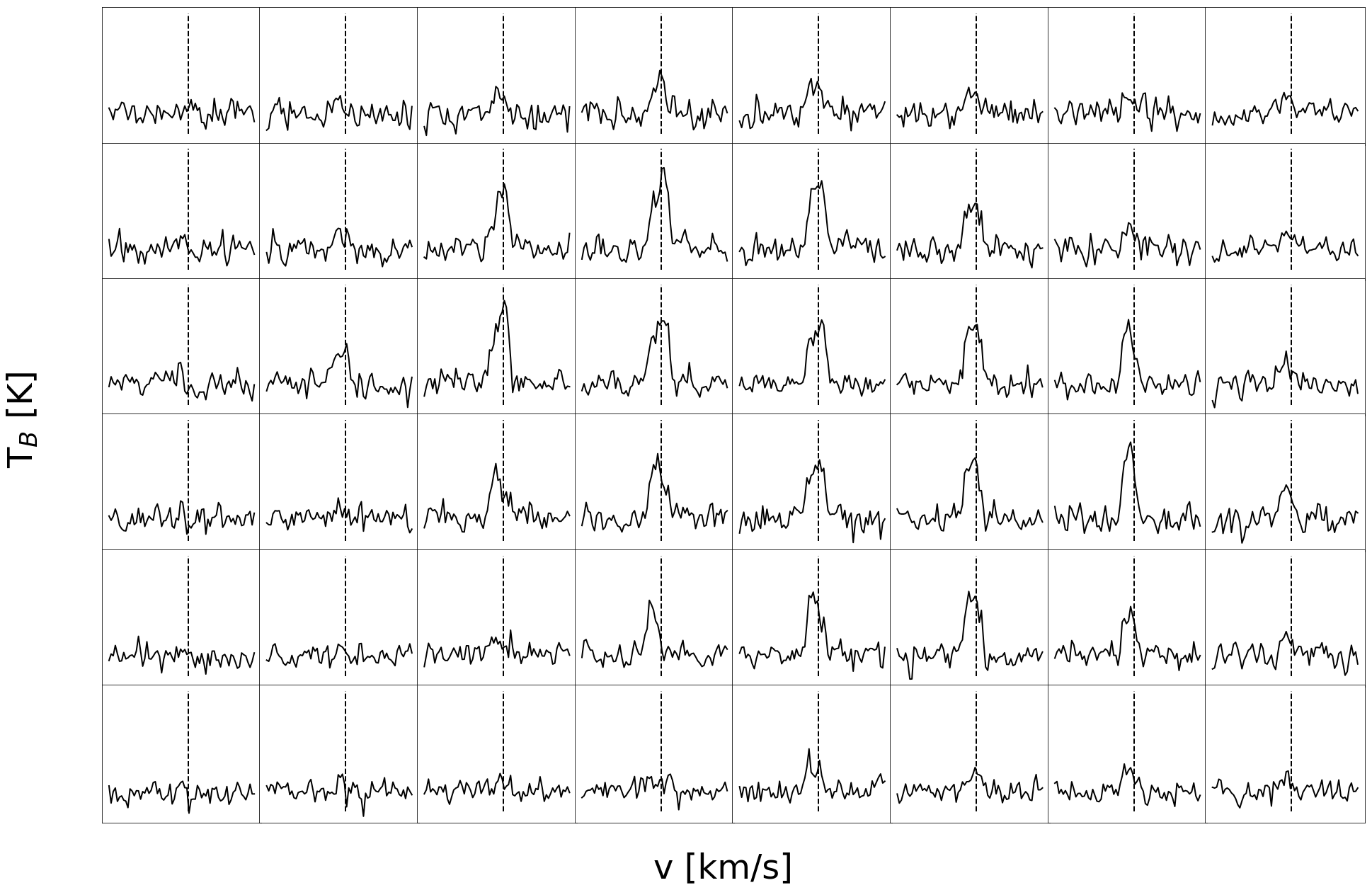}
    \includegraphics[trim=-10mm 0mm 0mm 5mm, width=8.5cm]{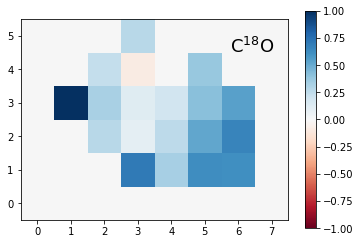}
    \caption{
    \textit{Top:}
    Moment~0 map of the $^{12}$CO J=2-1 emission with a grid showing the regions used to extract the spatially-resolved line profiles shown below. Red and blue contours show the HH~900 bipolar outflow (integrated CO J=2-1 emission from [--30,--11]~\kms\ and [--57,--36]~\kms, respectively). 
    \textit{Left:}
    Spatially-resolved line profiles with a dotted line indicating the $v_{\mathrm{LSR}}= -33.5$~\kms\ overplotted.  
    \textit{Right:} Maps of the line asymmetry quantified using the $A$ parameter from \citet{jackson2019}, see Section~\ref{sss:cold_kin}. 
    }
    \label{fig:line_profiles}
\end{figure*}

Unlike other studies, we spatially resolved the cold molecular gas and its kinematics.  
We examine line profiles as function of position in Figure~\ref{fig:line_profiles}. Profiles of each isotopologue are extracted from a rectangular grid where the width of each box is roughly twice the diameter of the beam. 
While single-peaked, most of the line profiles in Figure~\ref{fig:line_profiles} are not Gaussian. 
Pervasive line asymmetries that skew blue have been interpreted as evidence for infall \citep[e.g.,][]{walker1986,gregersen1997,mardones1997,lee1999,wu2003,fuller2005,reiter2011_kinematics}. 
To quantify the line asymmetries in the tadpole globule, we use the streamlined asymmetry analysis from \citet{jackson2019}. 
The asymmetry parameter $A$ is defined as:  
\begin{equation}
    A = \frac{I_{\mathrm{blue}} - I_{\mathrm{red}}}
             {I_{\mathrm{blue}} + I_{\mathrm{red}}}
\end{equation}\label{eq:a_param}
where $I_{\mathrm{blue}}$ and $I_{\mathrm{red}}$ are the integrated intensities   blueward and redward, respectively, of the $v_{\mathrm{LSR}}$.  
This parameter is independent of assumptions about the line shape and provides a simple quantitative assessment of the amount of flux in the line profiles at velocities bluer than the $v_{\mathrm{LSR}}$. 
For symmetric line profiles, $A=0$; blue-asymmetric profiles have positive values ($A>0$) while lines that skew red have negative values ($A<0$). 
We perform this spatially-resolved analysis for the J=2-1 transition of all three CO isotopologues where the line peak is detected with $\geq 6 \sigma$ significance. 
Maps of $A$ computed as a function of position across the tadpole head are shown in Figure~\ref{fig:line_profiles}.

Using this approach, line profiles appear to be blue overall, with redder line asymmetries around the redshifted side of the outflow. 
This is most clearly seen in $^{12}$CO although redder line profiles are seen in the same portion of the globule in the \tco\ and \cdo\ maps, even though outflow emission appears to be optically thin (see Paper~II). 
The extent to which the outflow influences the global kinematics of the globule is unclear. 
Given the relatively wide opening angle of the outflow ($\sim 50^{\circ}$), most of the energy and momentum may be deposited outside of the cloud. 
However, if the evolution of the outflow cavity is on-going, the outflow may continue to drive larger-scale motions within the globule. 
 
One additional source of uncertainty in this analysis is the $v_{\mathrm{LSR}}$ as none of the detected molecular lines have Gaussian profiles, although multiple tracers are consistent with $v_{\mathrm{LSR}} = -33.5$~\kms\ (Paper~II).

\begin{figure}
    \includegraphics[width=8.75cm]{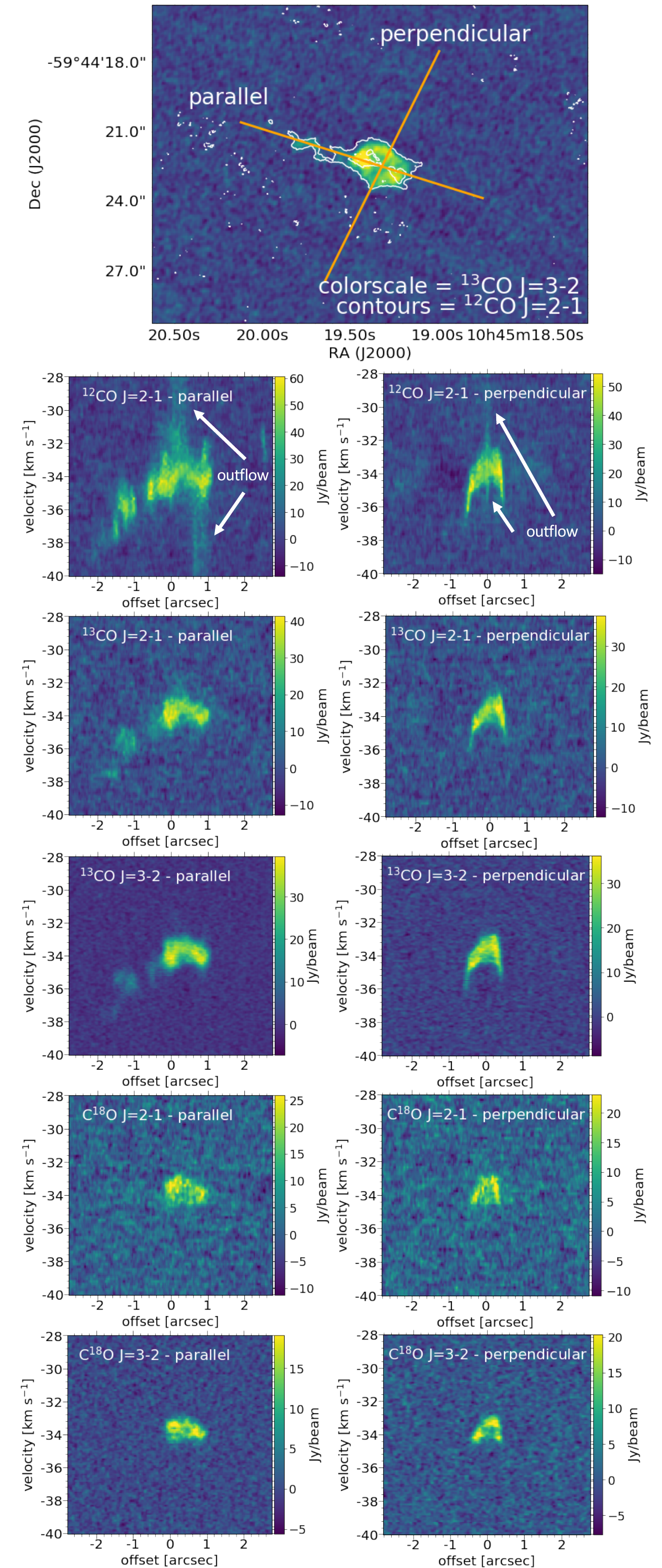}
    \caption{Position-velocity (P-V) diagrams showing gas velocities along cuts parallel and perpendicular to the outflow axis (\textit{top}). 
    Emission in the P-V diagrams is summed over slices $0.25\arcsec$ wide. 
    High-velocity gas in the $^{12}$CO diagrams traces the HH~900 molecular outflow. 
    }
    \label{fig:PV_diagrams}
\end{figure}

Position-velocity (P-V) diagrams provide another perspective on the spatially-resolved gas kinematics. P-V diagrams 
of all of the observed CO lines and isotopologues are shown in Figure~\ref{fig:PV_diagrams}. 
High-velocity features seen in both of the $^{12}$CO J=2-1 P-V diagrams trace the molecular outflow associated with HH~900. 
All diagrams show a C-shaped morphology in the globule head; this has been interpreted as infall in other sources \citep[e.g.,][]{keto2002,keto2006}. 

\subsection{Gravitational boundedness of the outer globule}
\label{sec:boundedness}
In Paper~I, we used MUSE to detect 
a photoevaporative wind emanating from the globule. In Paper~II, we presented evidence for heating of the surface layers (see also Section~\ref{ss:environ}). 
We know there is an ionised photoevaporative wind and warm molecular upper layer, but where is the wind actually launched? Is there a slow warm photodissociation region (PDR) flow before the photoionised flow traced by MUSE, or instead a contact discontinuity at the ionisation front?  Understanding the degree of boundedness of the warm outer layers of the globule interior to the photoionised wind could also constrain the mass (both in stars and gas) within the globule. 

The gravitational radius is that at which the thermal energy exceeds the gravitational energy \citep[e.g.][]{1994ApJ...428..654H}: 
\begin{equation}
    R_g = \frac{GM}{c_s^2},
\end{equation}
and material with some sound speed $c_s$ at a distance $R>R_g$ from a mass $M$ will be gravitationally unbound.

\begin{figure}
    \hspace{-0.3cm}
    \includegraphics[width=9.5cm]{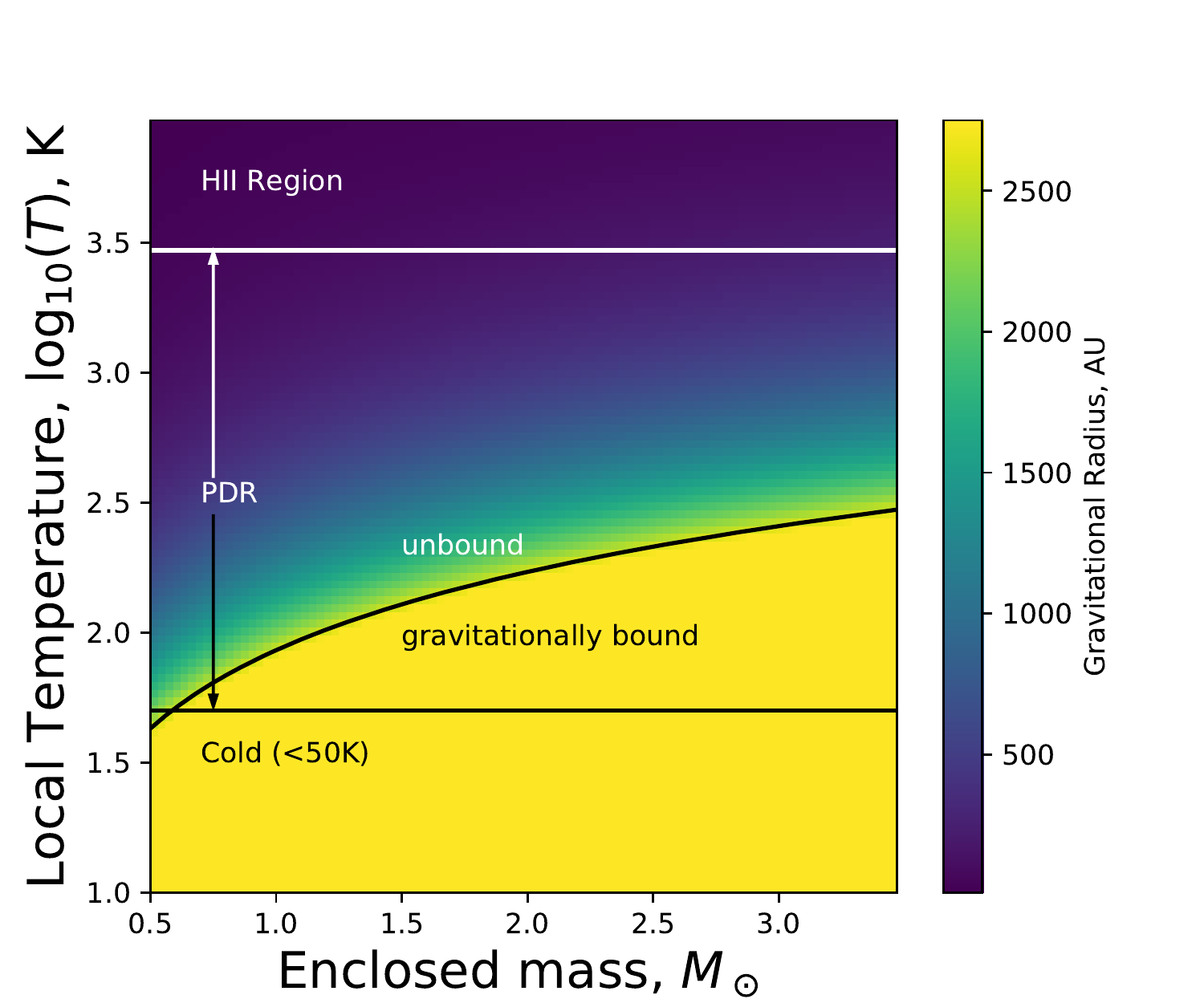}
    \caption{Each point on this plot represents the radius beyond which material at a given temperature will escape from some enclosed (star+globule) mass. The black contour denotes where the gravitational radius is the size of the globule (2500\,au). The horizontal lines separate cold gas, the photodissociation region and photoionised gas temperatures. Since we infer $60-100\,$K in the outer part of the glouble, we expect that outer molecular parts of the globule are bound. The wind will be driven in the warmer PDR ($T>300$\,K), which is expected to be thinly separated from the ionisation front in this case. }
    \label{fig:Rgrav}
\end{figure}

We plot the gravitational radius in terms of the globule-star mass and temperature in Figure \ref{fig:Rgrav}. A point on this plot represents material at some temperature, exterior to some amount of enclosed mass. The colourscale then represents the gravitational radius -- the distance at which the material of that temperature would be unbound from the enclosed mass. If the actual distance of material is less than the gravitational radius then it would be gravitationally bound. Similarly if the actual distance of material from the enclosed mass is greater than the gravitational radius, it is unbound. The black curve in Figure \ref{fig:Rgrav} is the contour where the gravitational radius is equal to the radius of the globule (2500\,au). The outer molecular layers will only be unbound if they have temperatures above this contour (and hence a gravitational radius interior to the globule outer edge). 

For a globule radius of 2500~au and interior mass from 0.5 to 2\,M$_\odot$ we find that the warm outer regions ($60-100$\,K) that we observed are most likely to be gravitationally bound. Material is only expected to be liberated in the warmer PDR ($T>\sim 200$\,K). The width of this warmer PDR zone interior to the ionisation front is unknown, but thin given the relative location of the MUSE emission (Paper~I).  
This is consistent with  
a relatively dense post-collapse globule that is not undergoing widespread loss of its outer layers. 
The interior kinematics of the globule are hence unlikely to be associated with strong bulk thermal expansion of the outer layers of the globule. 
A species such as CI may offer a probe of the inner $\sim300\,$K wind-launching region.

\subsection{Interpreting the position-velocity diagram morphology}
The spatially-resolved line profiles of the globule do not exhibit  non-symmetric double-peaked shapes characteristic of global infall or outflow \citep[e.g.][]{1996ApJ...465L.133M, 2012ApJ...750...64S}. This, coupled with the fact that there is an embedded YSO, suggests that this globule is in the post-RDI phase where it is undergoing less dramatic, relatively steady photoevaporation. 

The P-V diagrams made from cuts across the globule show more interesting and complicated kinematic structure. There is a characteristic C-shaped morphology that appears for cuts both perpendicular and parallel to the HH~900 jet+outflow system and appears in all of the CO isotopologues (Figure~\ref{fig:PV_diagrams}). This could possibly arise from ongoing infall (perhaps feeding the high accretion rate that is usually associated with jet launching) if the continuum optical depth were high enough to obscure any infall from the far side of the globule. 
For lower continuum optical depths, emission would trace a full loop in the P-V diagram. 
Note that the continuum opacity (not the line opacity) is critical to obscure emission from the far side of the globule, otherwise the far side will still be observed owing to the different velocity in the line, e.g. as a blue-asymmetry \citep[][]{2012ApJ...750...64S}.

To assess whether this is possible, we compute the column at which the continuum would be optically thick. That is, where
\begin{equation}
    \tau_\nu = \kappa_\nu N_{H2} \frac{2.7 m_H}{100} =   \kappa_\nu N_{i} \frac{2.7 m_H}{100 X_i} = 1
\end{equation}
for species $i$ with abundance $X_i$ and column $N_i$ where the opacity is per gram of dust. 
The factor of 2.7 is the mean mass per hydrogen molecule. 
We estimate the opacity using \cite{2003ApJ...598.1017D} silicates with a minimum grain size of 0.1~\micron, a maximum size of 1~mm, and a power law of distribution of $p=3.2$. This is a more evolved grain distribution than typically adopted in the ISM but  consistent with evidence for grain growth in the globule (see Paper~II) and will act to increase the opacity of the CO J=2-1 line relative to a less evolved dust population. 
The absorption opacity we adopt is $\kappa_{\textrm{abs}} \approx 2.5$\,cm$^2$\,g$^{-1}$. 
Assuming abundances by mass relative to H$_2$ of $8\times10^{-5}$,  $8.7\times10^{-6} $ and $1.7\times10^{-7}$ for $^{12}$CO, $^{13}$CO and C$^{18}$O respectively, we find that the continuum would be optically thick for CO columns of 
\begin{equation}
    N_{\textrm{12CO, crit}} = 7.1\times10^{20}\textrm{cm}^{-2}, 
\end{equation}
\begin{equation}
    N_{\textrm{13CO, crit}} = 2.4\times10^{19}\textrm{cm}^{-2}
\end{equation}
and
\begin{equation}
    N_{\textrm{C18O, crit}} = 1.5\times10^{18}\textrm{cm}^{-2}. 
\end{equation}
All of these critical columns exceed the column densities reported in Paper~II by roughly two orders of magnitude (well in excess of uncertainties on the relative abundances). The scattering opacity can further increase the extinction by a factor 5-10, but this is still insufficient to render the continuum optically thick. 
The C-shape morphology cannot therefore be due to some process happening over the entire globule surface with dust obscuring the far side of the globule. If the globule outer layers were all expanding or collapsing we should be able to see it happening on the far side, which would give a ring rather than a C-shape.

Other scenarios, including a molecular wind from the surface of the globule being blown toward the observer, are also difficult to reconcile with the data. 
Any such wind would have to be molecular in order to produce the C-shape in the P-V diagrams in Figure~\ref{fig:PV_diagrams}. 
This is only possible if the outer molecular layers are unbound, but this is not expected based on the analysis in Section~\ref{sec:boundedness}. 
Only warmer (neutral and/or ionised) gas is unbound and may be launched in such a wind from the globule surface. 
Our simple analysis does not take into account other external influences on the globule (i.e.\ stellar winds from the nearby O-type stars) that may lead to more complex bulk dynamics \citep[e.g.,][]{bally1995}. 
We cannot ascertain exactly what is responsible for the C-shape, but can say that it is not 
an expanding or collapsing spherical globule.

\section{Bulk kinematics of the globule}\label{s:bulk_kin}

\subsection{The unusual velocity of the tadpole} 
In Paper~II, we measured a $v_{\mathrm{LSR}} = -33.5$~\kms\ for the \tyso\ and globule, which is $\gtrsim 10$~\kms\ 
bluer than the typical velocity of the molecular gas in the direction of Carina \citep[$\sim -20$~\kms,][]{rebolledo2016}. 
Gas in the tadpole tail is further blueshifted relative to the globule velocity, with emission extending up to $\sim -37.5$~\kms\ (see Figure~\ref{fig:PV_diagrams}). 
Emission from the system is spatially and spectrally continuous, with a smooth increase in velocities from the head of the tadpole through the tail, suggesting that the features are physically associated.

\subsection{Why is the tadpole travelling so fast?}

The tapdole has a blue-shifted bulk velocity of 10\,km\,s$^{-1}$ relative to the local standard of rest, which is far in excess of the $\sim1$\,km\,s$^{-1}$ turbulent velocity dispersion in the nearby molecular gas \citep{rebolledo2016}. We now explore whether this could be a consequence of the globule formation mechanism, or a consequence of the stellar irradiation driven rocket effect.

\subsubsection{A globule born at high speed?}\label{sss:born}
There are two primary classes of globule formation within star forming regions. In the first class, globules are associated with supernova events and have very high velocities. In the Crab Nebula, globule velocities can be as high as 60--1600\,km\,s$^{-1}$ \citep{2017A&A...599A.110G} and in the Rosette the expansion velocity of the shell, pillars and globules are all $\sim22$\,km\,s$^{-1}$ \citep{2013A&A...555A..57G}. 

A supernova event could be responsible for the high velocity of this tadpole globule. In Carina, there is a candidate neutron star identified by \cite{hamaguchi2009} at a projected distance from the globule that is similar to the projected distance of $\eta$ Car ($\sim3.5\,$pc; see Figure~\ref{fig:hh900_intro}). However, beyond this neutron star we have very limited understanding of when and how a supernova may have gone off in this region. Furthermore, the mechanism for producing globules from supernova events is also quite uncertain. We therefore note that a supernova is a possibility, but focus our attention on the second class of mechanism, which involves the interplay between ionising radiation and the surrounding ISM.

To discuss how ionising radiation can contribute to producing globules, we need to briefly summarise the broad behaviour of H\,\textsc{ii} regions. For the simplest case of a star emitting $N_{ly}$ ionising photons per second into a uniform medium of number density $n$, gas is quickly ionised within a sphere of ``Str\"{o}mgren'' radius 
\begin{equation}
    R_s = \left(\frac{3N_{ly}}{4\pi \alpha_B n^2}\right)^{1/3}
\end{equation}
where $\alpha_B$ is the case B recombination coefficient for hydrogen. This Str\"{o}mgren sphere is overpressured relative to the ambient ISM and so expands in time as
\begin{equation}
    R_I = r_s\left(1+\frac{7\sqrt{4}c_it}{4\sqrt{3}R_s}\right)^{4/7}
    \label{equn:HII}
\end{equation}
in the limit that the H\,\textsc{ii} region number density does not drop to so low a level that the ionised and neutral pressures become comparable \citep{2006ApJ...646..240H, 2015MNRAS.453.1324B, 2018MNRAS.479.2016W}. The intial speed of this expanding region is hence approximately the ionised gas sound speed ($c_i=11.5$\,km\,s$^{-1}$ for $8\times10^3$\,K gas), but decays as
\begin{equation}
    \dot{r}_I = \frac{\sqrt{4}}{\sqrt{3}}c_i\left(1+\frac{7\sqrt{4}c_it}{4\sqrt{3}r_s}\right)^{-3/7}. 
    \label{equn:HIIvel}
\end{equation}

With this in mind, common frameworks for globule production associated with ionising radiation are:

\begin{figure*}
    \centering
    \includegraphics[width=14cm]{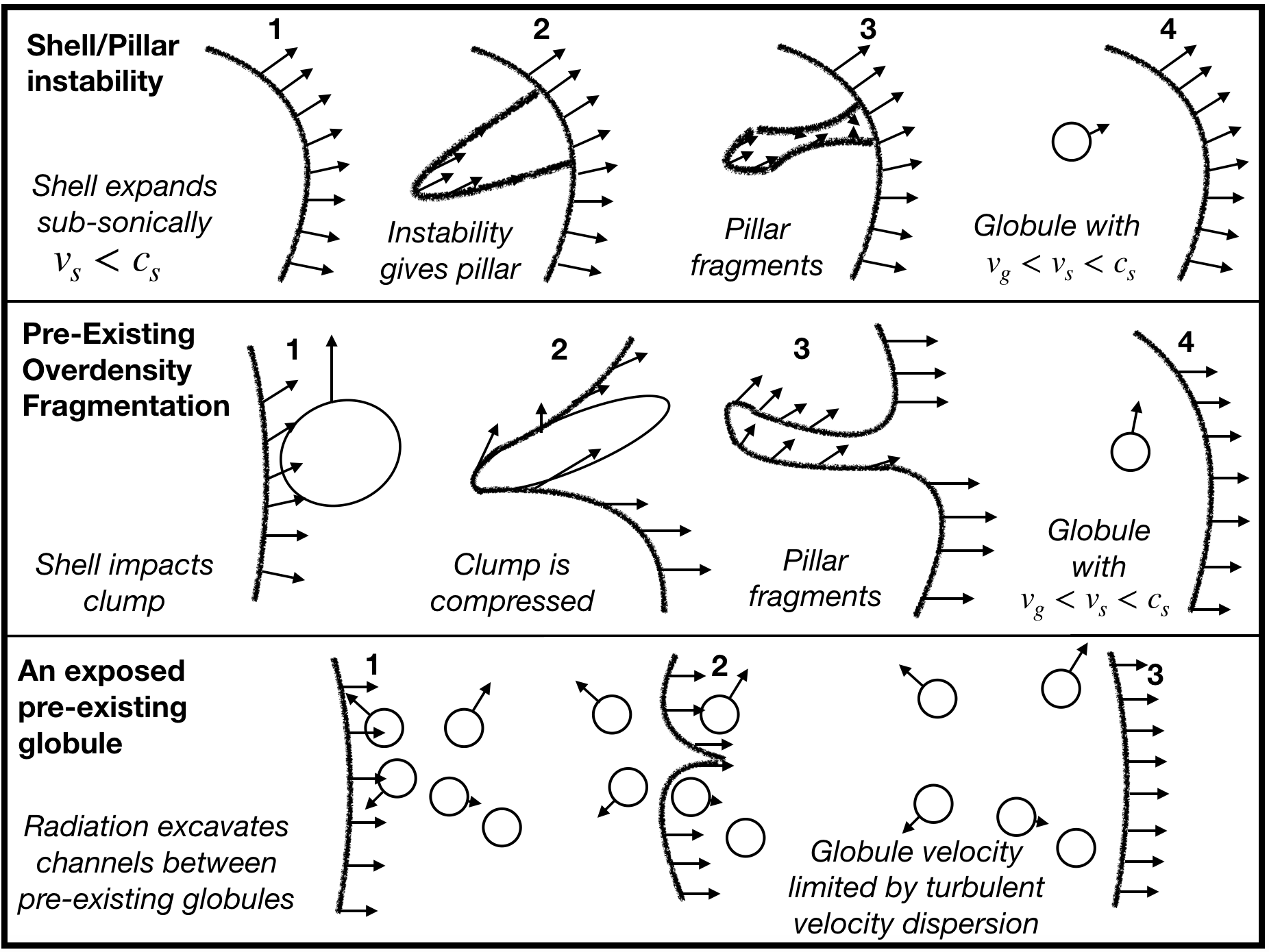}
    \caption{Cartoons of the various formation mechanisms for globules. 
    \textit{Top:} In shell/pillar instability an expanding shock/ionisation front becomes unstable. This results in a pillar that itself fragments to form globules. Since the pillar propagates more slowly than the shell, the globule velocity is initially slower than the shell. 
    \textit{Middle:} In the second mechanism, a shell can impact a pre-existing clump or filament. In this case a pillar forms and fragments, but the globule velocity may also inherit kinematic characteristics of the precursor cloud. 
    \textit{Bottom:} In the third mechanism the globules already reside in the neutral ISM and are exposed when the region is irradiated. These will have velocities limited by the turbulent velocity dispersion. None of these mechanisms can give velocities faster than the sound speed and hence (typically) the expanding shell. }
    \label{fig:cartoons_form}
\end{figure*}

\begin{figure*}
    \centering
    \includegraphics[width=14cm]{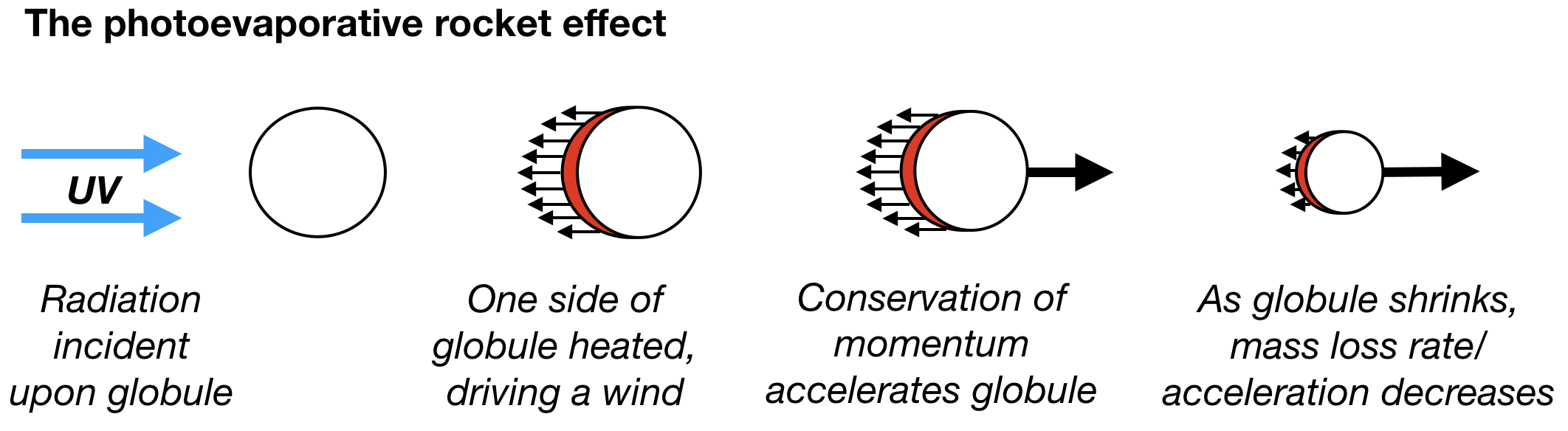}
    \caption{An illustration of the photoevaporative rocket effect, wherein one side of a globule is irradiated leading to a wind, which causes the globule to propagate in the opposite direction. This can accelerate globules to speeds in excess of the sound speed (see Figure \ref{fig:rocketModel}). }
    \label{fig:cartoons_rocket}
\end{figure*}

\begin{enumerate}
    \item \textbf{Shell/pillar instability:} In this model the expanding H\,\textsc{ii} region sweeps up a shell that becomes dynamically unstable to form pillars \citep[also referred to as ``elephant trunks'', ][]{1954ApJ...120...18F, 1996ApJ...469..171G}. Forming from the shell, these pillars move outward from the ionising source with velocities comparable to (slightly lower than) the H\,\textsc{ii} region expansion speed. Globules result from fragmentation of these pillars. In the absence of further acceleration, the globules have an outward velocity lower than the H\,\textsc{ii} region expansion speed at the time they detached from the shell. 
    
    The initial H\,\textsc{ii} region expansion speed is similar to the current speed of the tadpole ($\sim 10$~\kms), but the expansion speed halves in only a few hundred thousand years. Pillars/globules formed this way have an outward speed that is lower than the shell of the H~{\sc ii} region, making this formation mechanism an unlikely explanation for the high tadpole velocity. \\

    \item\textbf{Pre-existing overdensity fragmentation:} This mechanism differs from the first in that pillars form from  pre-existing overdensities (e.g., filaments) rather than as a  result of instability in the shell \citep[e.g.][]{2010MNRAS.403..714M, dale2013}. As in the first case, globules form from fragmentation of this structure. 
    
    In this framework, the pillar and resulting globule may have a systematically different velocity compared to the expansion of the H\,\textsc{ii} region, but differences are constrained by the turbulent velocity dispersion of the region. Therefore, this mode of globule formation cannot explain the tadpole velocity, which is in excess of the turbulent velocity dispersion. \\

    \item\textbf{An exposed pre-existing globule:} In this scenario, globules already exist in the turbulent neutral ISM. By virtue of being over dense, they are more resilient in the face of the radiation field that ionises the ambient gas \citep[e.g.][]{2012A&A...546A..33T, 2013A&A...560A..19T}. 
    
    This mechanism also requires that the globule motions are within the turbulent velocity dispersion and so cannot explain the observed tadpole velocity. 
\end{enumerate}
A schematic summary of these different formation processes is given in Figure \ref{fig:cartoons_form}. In summary, none of these globule formation mechanisms can explain the observed tadpole velocity. Unless the high globule velocity resulted from a supernova, this implies that the globule must have been subsequently accelerated.

\subsubsection{A rocket-driven tadpole?}
\label{sec:rocket}
Since the tadpole seems unlikely to have formed with its current velocity, it must have been accelerated (to some extent at least) at a later time, most likely through the photoevaporative rocket-effect \citep[e.g.][]{1998A&A...331..335M}. 
In this mechanism a flow is driven from the irradiated side of the globule. 
Momentum is conserved, so the momentum flux from the evaporating material ejected from the globule imparts a force of equal magnitude that accelerates the globule in the opposite direction, as illustrated in Figure \ref{fig:cartoons_rocket}. 
Globules are indeed accelerated to high speeds in simulations of photoionisation feedback in star forming regions \citep[e.g.][]{2009MNRAS.398..157H}. 

\begin{figure*}
    \centering
    \includegraphics[width=18.5cm]{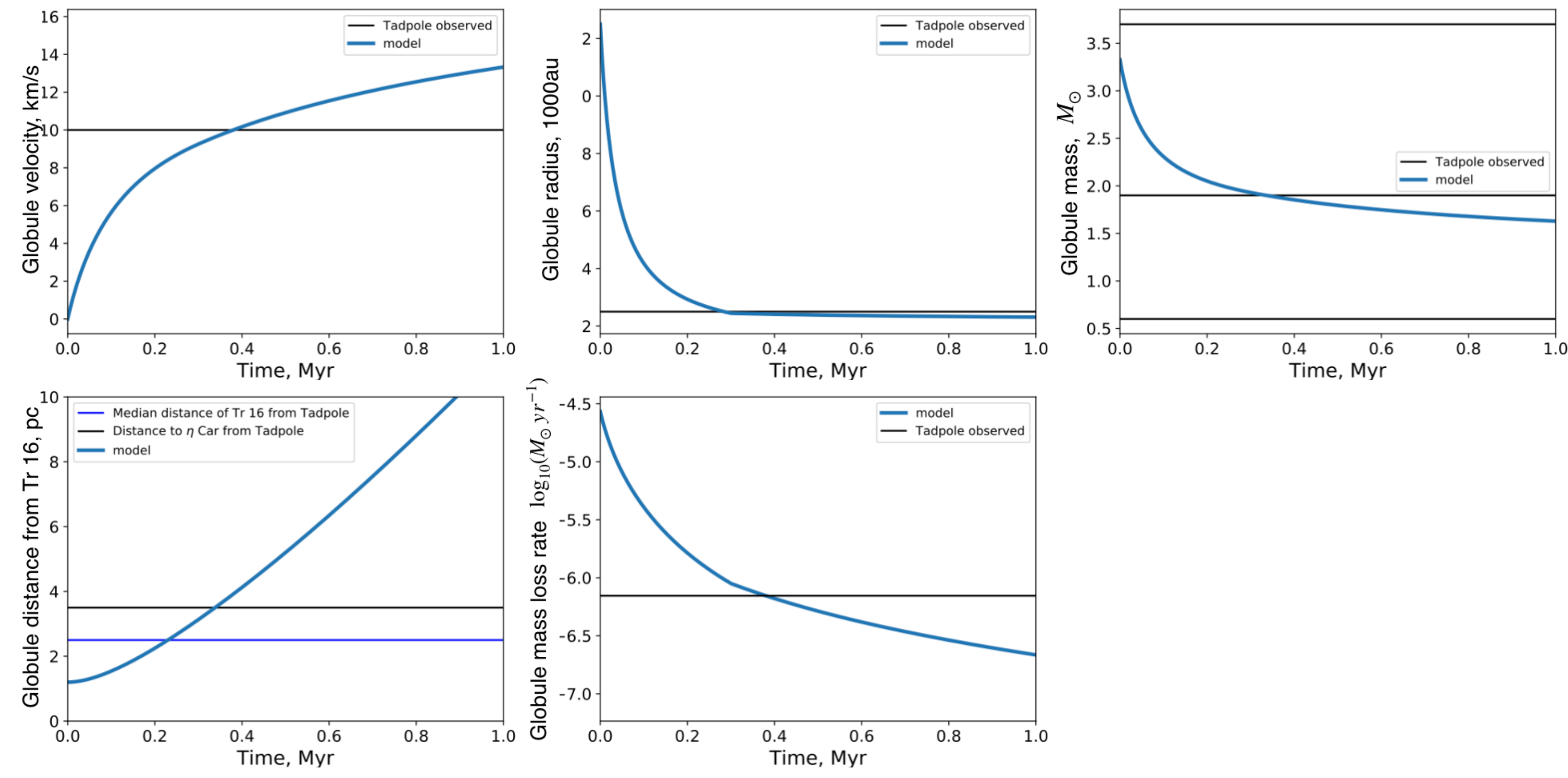}
    \vspace{-0.5cm}
    \caption{Semi-analytic models of the evolution of a globule that is being subjected to radiatively-driven implosion and the photoevaporative rocket effect. Panels are the globule velocity, radius, mass, distance from the ionising source and mass loss rate from left-to-right, top-to-bottom. Observed values for the tadpole studied here are marked using the horizontal black lines. The model is for a globule with initial mass of $3.3\,M_\odot$, initial radius of 12500\,au at a distance of 1.2\,pc from a $10^{50}$ ionising photon per second source. These simple models are not expected to exactly fit all parameters of the globule (which are degenerate), but do demonstrate that a radiatively-driven implosion plus photoevaporative rocket effect scenario can approximately retrieve all of the observed parameters at around 0.4\,Myr. }
    \label{fig:rocketModel}
\end{figure*}

We make our own basic assessment of the motion of the tadpole under the rocket effect using a simple semi-analytic approach. We consider the equation of motion of an evaporating clump of mass $M_c$, radius $R_c$ with evaporative outflow speed $v_p$ at a distance $D$ from an ionising source. Motion away from the ionising source is positive (and motion toward is negative). The equation of motion is given by 
\begin{equation}
	M_c\ddot{D} + \dot{M_c}\left[\dot{D} - v_p\right] = 0 
	\label{EOM_reduced}
\end{equation}
where the sign of the photoevaporative outflow speed accounts for the fact that it is in the negative direction. 
Here we have ignored the second order effects of gravity from the stellar cluster, ram pressure, radiation pressure, and the stellar wind. 
We use a mass-loss rate $\dot{M_c}$ for an evaporating globule from \cite{1998A&A...331..335M}
\begin{equation} 
	\dot{M_c} = - F' \mu m_H \pi \left(\frac{\pi R_c}{2}\right)^{2}
	\label{massloss}
\end{equation}
where 
\begin{equation}
	F' = \frac{F_o}{(1+\frac{\alpha_BF_oR_c}{3c_i^2})^{1/2}}
\end{equation}
and $F_o$ is the geometrically diluted ionising flux at the clump. We assume that the photoevaporative outflow is launched at the sound speed of ionised gas.

For the evolution of the globule radius with mass we take a two-component approach to emulate the process of more dramatic RDI followed by a phase of steady evaporation of a collapsed globule. From the models of \cite{2009MNRAS.398..157H}, the early phase of compression leads to a factor $\sim 2$ reduction in neutral mass and factor $\sim 8$ reduction in radius in the first 0.8\,Myr. We therefore initially evolve the radius as  $R_g\propto M_g^3$, and then after 0.3\,Myr assume that the evaporation is slow and steady and the globule density remains constant (i.e. $R_g\propto M_g^{1/3}$).

With the above framework we find that an initial globule mass of $\sim3.3\,$M$_\odot$ and radius of 12500\,au starting with zero velocity at a distance of 1.2\,pc from the ionising source of $10^{50}$\,s$^{-1}$ can broadly reproduce the mass, radius, velocity, position and mass-loss rate of the globule, as shown in Figure~\ref{fig:rocketModel}. These initial conditions also have an ionising flux $ \Phi/\Phi_{\mathrm{crit}} = 0.35$ (see equation \ref{equn:phicrit}) so star formation could still take place through RDI. 
These parameters are degenerate with others that would satisfy the currently observed conditions and we do not claim to have a model for the evolutionary history of the globule. 
Nevertheless, the rocket effect provides a plausible explanation for the high velocity of the tadpole whereas none of the formation mechanisms described in Section~\ref{sss:born} can explain the kinematics of the globule on their own.

\section{Triggered star formation?}\label{s:triggered}

\subsection{Bending of the HH~900 jet}
\begin{figure}
    \centering
    \includegraphics[width=8cm]{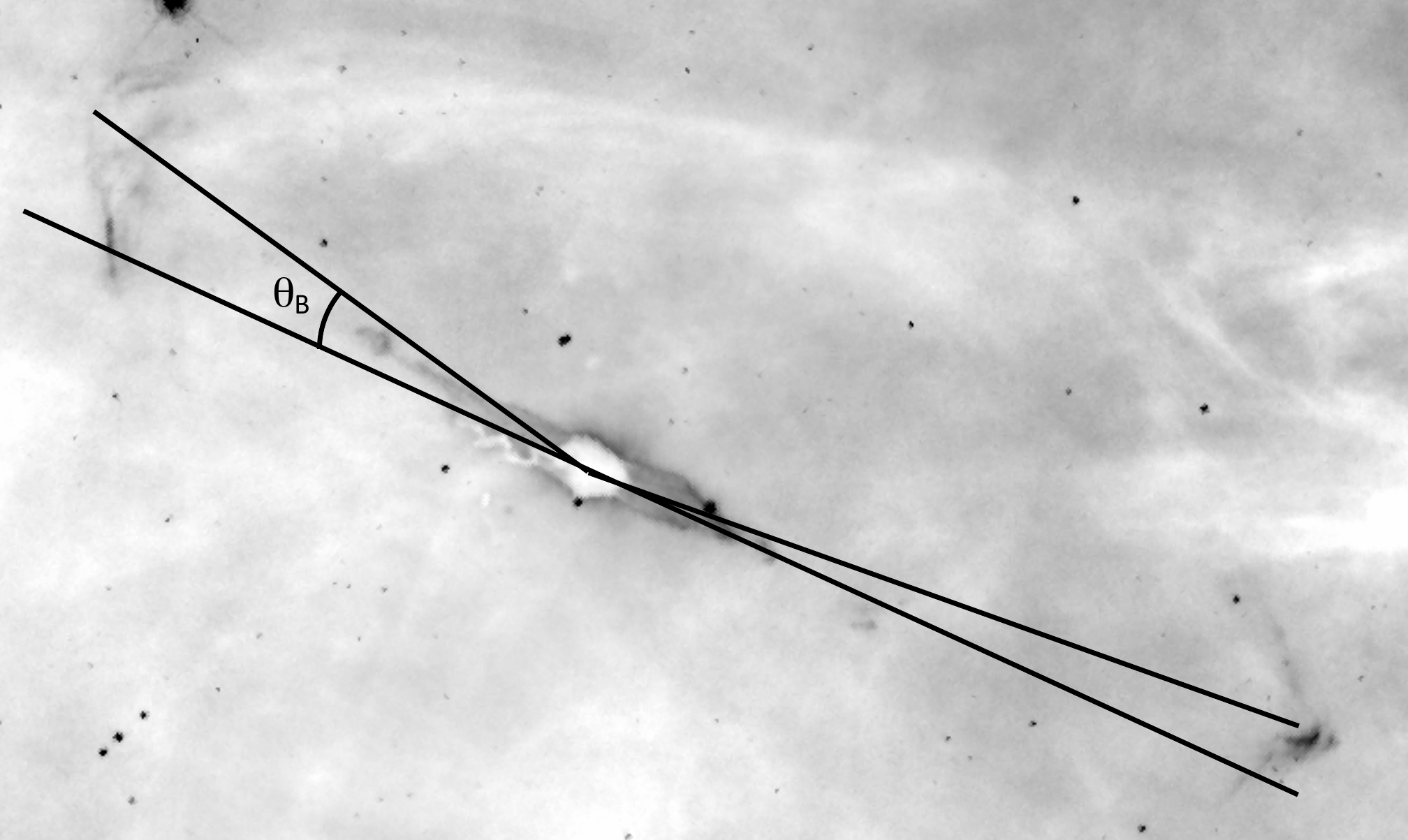}
    \caption{\emph{HST} image of the HH~900 jet+outflow systems with lines overplotted that show a straight line through the \tyso\ (bottom) and the vectors from the \tyso\ that intersect the bow shocks that are offset from this line by an angle $\theta_B$. Unlike other jets in H~{\sc ii} regions, HH~900 bends \emph{toward} the ionizing sources. }
    \label{fig:jetBending}
\end{figure}

Protostellar jets in stellar clusters can be deformed by the environment (e.g., via stellar winds and/or the rocket effect) and exhibit a range of  shapes \citep[e.g.,][]{bally2006}. In the case of HH~900, the jet bends toward the main ionising sources, as illustrated in Figure \ref{fig:jetBending}. The terminal bow shocks are offset from the continuous inner jet by $\sim 2.9\arcsec$.
\citet{reiter2015_hh900} argued that this might be due to the photoevaporative flow from the dust wall behind the jet, effectively providing stronger ram pressure. 

Jet bending could also result if the globule and the jet-driving YSO embedded within it are moving fast enough to be noticeably displaced over the jet traversal time.  
Imagine a small knot of material in the jet. 
In the reference frame of the YSO, the bipolar jet (including the knot) is launched in a straight line, perpendicular to the disk. 
Once set in motion, the jet knot continues to travel in a straight line (neglecting any interaction with ambient material in the globule). 
However, the globule+YSO are moving perpendicular to the jet axis, so at a later time, the knot appears to be offset from current jet axis by an angle $\theta_B$ (see Figure~\ref{fig:jetBending}). 
In this case, if the jet velocity is $v_{jet}$, the jet length is $l_{jet}$, and the jet-driving source is moving at $v_{*}$, the origin of the jet will have moved $\Delta = l_{jet}v_*/v_{jet}$.  

To determine if the observed jet bending is consistent with the outward motion of a globule accelerated by the rocket effect, we first estimate the traversal time of the globule. 
The HH~900 bow shocks are located $\sim 23\arcsec$ from the \tyso, corresponding to $\sim 0.25$~pc for a distance of 2.3~kpc. 
Proper motions indicate that the transverse velocities of the bow shocks are $\sim 60$~km~s$^{-1}$ \citep{reiter2015_hh900}, leading to a 
traversal time $\sim 4200$~yr. 
If the tadpole globule has maintained a steady velocity of 10~km~s$^{-1}$, it will have moved the observed $\sim 2.9\arcsec$ in $\sim 3200$~yr. 
This rough estimate does not take into account inclination effects, nor any velocity evolution of the jet (i.e., material in the bow shocks may have had a faster velocity in the past leading to a younger dynamical age). 
Nevertheless, the similarity of the expected and observed displacement is suggestive, indicating that motions of the jet-driving source can account for the jet bending.

The above raises the interesting notion that if the globule has been accelerated (see Section~\ref{sec:rocket}) and the jet-driving YSO is still approximately co-moving \textit{then star formation must have initiated once the globule was already up to speed}. 
If the star formed first and then the globule was accelerated, the star would have been left behind and there would be no jet bending (at least due to globule motion). If the jet bending is due to the stellar motion, and this motion was set by the radiation driven acceleration of a globule, this is suggestive (but not conclusive proof) of triggered star formation. 

\subsection{An unusual mode of star formation? }

The ALMA data presented in Paper~II revealed the embedded YSO that drives the HH~900 jet+outflow for the first time. 
It appears to be a single star surrounded by at least 1~\Msun\ of dust, roughly 50\% of the estimated mass of the globule.  
While we do not have a direct estimate of the mass of the YSO (see Paper~II), 
the high mass-loss rate of the HH~900 jet+outflow system is among the highest of the known jets in Carina \citep{smith2010} leading \citet{reiter2015_hh900} to suggest a driving source mass $\gtrsim 2$~\Msun. 
This suggests that the \tyso\ will be at least an A-type star, among the earlier spectral types where the companion frequency is $\sim 100$\% \citep{duchene2013}. 
The separation distribution for A-type stars appears to peak around $\sim 100$~au \citep{duchene2013}, similar to the spatial resolution of our ALMA data (see Paper~II).  
A possible a second peak of close-separation sources around A-type stars \citep[see, e.g.,][]{murphy2018} would be unresolved with our ALMA data.

The Jeans length, $\lambda_J \propto (T/\rho)^{1/2}$, indicates that lower density gas will be unstable at lower temperatures. 
Regions like Carina that have significant radiative feedback may have higher temperature gas that will be more stable against fragmentation. 
In a recent study of two regions in Carina subject to different amounts of feedback, \citet{rebolledo2020} find that the warmer, more intensely irradiated region hosts fewer but higher mass cores than the more quiescent region. 
If the \tyso\ is truly a single star despite the high companion fraction for stars of similar mass, then higher gas temperatures in the tadpole-shaped globule may have suppressed fragmentation.

\section{Conclusions}

In this paper, the third in a series, we consider the history and fate of a small tadpole-shaped globule located in the Carina Nebula. 
We compare results from optical integral field spectroscopy with MUSE (Paper~I) and millimeter maps from ALMA (Paper~II) 
with theoretical models to estimate how the environment affects the observed properties of the globule.  

Observations provide clear evidence that the environment continues to affect the globule, for example, by heating gas on the surface of the globule. 
We make multiple estimates of the past and present impact of feedback and how it may have affected the evolution of the globule. 
The current pressure of the ionization front does not dominate over neutral gas pressure in the high-density globule.  
Internal gas kinematics 
are ambiguous, but line profiles and P-V diagrams do not 
display traditional signatures of infall. 
Cold material seen with ALMA is gravitationally bound, consistent with a dense globule that is not experiencing widespread loss of its outer layers. 
Together, this suggests that the globule is in a post-collapse phase. 

Within the context of the Carina star-forming complex, the tadpole-shaped globule is blueshifted $\sim 10$~\kms\ relative to the systemic velocity of Carina. 
This velocity difference is too high to be explained by turbulence, suggesting that the globule was accelerated away from the high-mass stars in Carina by the photoevaporative rocket effect. 
The HH~900 jet+outflow system bends \emph{toward} the ionizing sources and we find that jet bending is consistent with the expected displacement if the jet-driving source and its natal globule have been accelerated by the rocket effect. 
This outward motion of star+globule is only possible if the star formed after globule acceleration, indicating that the formation of the YSO may have been triggered by RDI. 

We find that these features can be explained by a scenario in which the globule was originally much larger and closer to the ionizing sources. 
Radiatively-driven implosion accelerated the globule to the high observed speeds through the photoevaporative rocket effect and triggered the formation of the star that now drives the HH~900 jet+outflow. 
In this picture, the globule may now be in a quasi-steady state, post-collapse phase.

Finally, the data are consistent with a single star forming in the globule, despite the high companion frequency observed for stars of similarly high mass. 
If the YSO is in fact a single star, this would be unusual, suggesting that feedback may have heated the gas and thus suppressed fragmentation in the globule.

\section*{Acknowledgements}
We thank the referee, Dr G\"{o}sta Gahm, for a timely and thoughtful report. 
The authors wish to thank Joseph C. Mottram and David Rebolledo.  
This project has received funding from the European Union's Horizon 2020 research and innovation programme under the Marie Sk\'{l}odowska-Curie grant agreement No. 665593 awarded to the Science and Technology Facilities Council. 
T.J.H is funded by a Royal Society Dorothy Hodgkin Fellowship. 
A.F.M. is funded by a NASA Hubble Fellowship. 
G.G. acknowledges support from CONICYT project AFB-170002. 
This paper is based on data obtained with ESO telescopes at the Paranal Observatory under programme ID 0101.C-0391(A). 
This paper makes use of the following ALMA data: ADS/JAO.ALMA\#2016.1.01537.S. ALMA is a partnership of ESO (representing its member states), NSF (USA) and NINS (Japan), together with NRC (Canada) and NSC and ASIAA (Taiwan) and KASI (Republic of Korea), in cooperation with the Republic of Chile. The Joint ALMA Observatory is operated by ESO, AUI/NRAO and NAOJ.
This research made use of Astropy,\footnote{http://www.astropy.org} a community-developed core Python package for Astronomy \citep{astropy:2013, astropy:2018}.
This research made use of APLpy, an open-source plotting package for Python \citep{robitaille2012}.

\section*{Data Availability Statement}
The data underlying this article are available from the ALMA Science Archive at https://almascience.nrao.edu/asax/ and the ESO Science Archive Facility at http://archive.eso.org/.





\bibliographystyle{mnras}
\bibliography{bibliography_mrr} 



\bsp	
\label{lastpage}
\end{document}